\documentclass[letter]{aa}
\usepackage[varg]{txfonts}
 
\usepackage{amsmath}
\usepackage{amsfonts}
\usepackage{amssymb}
\usepackage{graphicx}
\usepackage{fourier}

\usepackage{natbib}
\bibpunct{(}{)}{;}{a}{}{,} 

\usepackage{booktabs} 

\usepackage{makecell}
\usepackage{txfonts}
\usepackage{graphicx,natbib,url,twoopt}
\usepackage{lipsum}
\usepackage{multicol}
\usepackage{float}

\usepackage{stackengine}
\usepackage{threeparttable}
\usepackage{subfig}
\usepackage{longtable}
\usepackage{listings}
\usepackage{placeins}

\usepackage{hyperref}

\begin{document}

\title{ARDENT: A Python package for fast dynamical detection limits with radial velocities}

\author{M. Stalport\inst{\ref{i:Liege_STAR}, \ref{i:Liege_EXOTIC}} 
\and M. Cretignier \inst{\ref{i:Oxford}}
\and L. Naponiello \inst{\ref{i:Torino_INAF}}
\and V. Van Grootel \inst{\ref{i:Liege_STAR}}
}

\institute{Space sciences, Technologies and Astrophysics Research (STAR) Institute, Université de Liège, Allée du 6 Août 19C, 4000 Liège, Belgium \label{i:Liege_STAR} 
\and Astrobiology Research Unit, Universit\'e de Li\`ege, All\'ee du 6 Août
19C, B-4000 Liège, Belgium \label{i:Liege_EXOTIC}
\and Department of Physics, University of Oxford, OX13RH Oxford, UK \label{i:Oxford} 
\and INAF - Osservatorio Astrofisico di Torino, via Osservatorio 20, 10025 Pino Torinese, Italy \label{i:Torino_INAF}
}

\date{Received -- / Accepted --}

\abstract
{The architecture of planetary systems is a key piece of information to our understanding of their formation and evolution. This information also allows us to place the Solar System in the exoplanet context. An important example is the impact of outer giant planets on the formation of inner super-Earths and sub-Neptunes. Radial velocity (RV) surveys aim at drawing statistical insights into the (anti-)correlations between giants and inner small planets, which remain unclear. These surveys are limited by the completeness of the systems, namely, the sensitivity of the data to planet detections. Here, we show that we can improve the completeness by accounting for orbital stability. 
We introduce the Algorithm for the Refinement of DEtection limits via N-body stability Threshold (ARDENT), an open-source Python package for detection limits that include the stability constraint. The code computes the classic data-driven detection limits, along with the dynamical limits via both analytical and numerical stability criteria. We present the code strategy and illustrate its performance on TOI-1736 using published SOPHIE RVs. This system contains an eccentric cold giant on a 570-day orbit and an inner sub-Neptune on a 7-day orbit. We demonstrate that no additional planet can exist in this system beyond 150 days due to the gravitational influence of the giant. This outcome allows us to significantly refine the system completeness and also carries implications for RV follow-ups. ARDENT is user-friendly and can be employed across a wide variety of systems to refine our understanding of their architecture.}


\maketitle

\nolinenumbers

\section{Introduction} \label{Sect:Intro} 
An important piece in the puzzle of planetary system formation and evolution is the absence of short-period super-Earths and sub-Neptunes in the Solar System. In recent decades, these planets have been revealed as very common planetary types in our Galaxy \citep[e.g.][]{Zhu2018}. The giants Jupiter and Saturn likely played a significant role in shaping the inner Solar System, and might constitute the root cause for the absence of super-Earths. This proposition can be further investigated in the exoplanet context, exploring the correlation between outer cold Jupiters (CJs) and short-period small planets (SPs) defined with 1 $< M_p <$ 20 M$_{\oplus}$ and P $\lesssim$ 100 d. 

Large radial velocity (RV) surveys have investigated the CJ-SP connection. While  a positive correlation was initially reported among CJs and SPs \citep[e.g.][]{Zhu2018}, this result was later challenged \citep{Barbato2018, Bonomo2023}. Recent studies analysing larger samples of systems suggest a tentative dependency of this correlation with the stellar host metallicity \citep[e.g.][]{Bryan2024, Bonomo2025}. To date, these studies face two major limitations: small number statistics and limited systems completeness. The former can only be alleviated in the near future by discovering many new planets, employing, for instance,  \textit{Gaia} Data Release 4 (DR4) for long-period giant planets and PLATO for new systems of SPs. The completeness, on the other side, is limited by the precision and sampling of the RV surveys. A deeper completeness increases the precision in occurrence rates, which is crucial to reveal potential correlations between CJs and SPs. Large observational efforts paired with novel techniques of stellar activity mitigation would be required to improve completeness. 

While this observational effort is ongoing, there is a straightforward way to tighten the RV detection limits within a given dataset, which involves the study of orbital stability. Based on the hypothesis that observed systems cannot have strongly unstable orbits (the probability of detecting a system transitioning to instability is very small), a planet can be considered detected if it fulfils two essential conditions: the signal in the data is statistically significant and it does not compromise the orbital stability of the system. Some works have included this dynamical constraint a posteriori on top of the data-driven detection limits, using simplified stability criteria \citep[e.g.][]{Bonomo2017, Barbato2023}. Nevertheless, there is currently no systematic inclusion of the stability constraint on the derivation of detection limits  for  CJ-SP studies or in general exoplanet contexts. Therefore, gaining access to an easy-to-use tool that derives such constraints is timely. 

In this Letter, we introduce the Algorithm for the Refinement of DEtection limits via N-body stability Threshold (ARDENT). It is an open-source Python package for the computation of dynamical detection limits; namely, detection limits that combine observational and dynamical constraints. Their computation is straightforward and can be applied to a wide range of systems. 
In Sect. \ref{Sect:Code}, we introduce the code and describe how the stability constraint is added to the data-driven detection limits. In Sect. \ref{Sect:TOI-1736}, we illustrate ARDENT based on TOI-1736. This is a two-planet system composed of a 7-day SP and 570-day CJ on an eccentric orbit. From this analysis, we derive the dynamical detection limits of TOI-1736, and sum up the advantages that ARDENT offers in Sect. \ref{Sect:Conclusion}

\section{ARDENT} \label{Sect:Code}
\subsection{General strategy}
The completeness of planetary systems from RV surveys is quantified via mass detection limits. Commonly, these are computed from injection-recovery tests with a various sample of hypothetical planets \citep[e.g.][]{Barbato2018, Rosenthal2021, Bryan2024}. The mass above which a certain percentage of injected planets can be recovered (e.g. 95$\%$) defines the 95$\%$ detection limits. Above these limits, we can firmly rule out the existence of additional planets within our threshold confidence level. In this work, we name these limits the data-driven detection limits, since they are directly related to the data at hand. 

However, these limits are incomplete because they do not validate the orbital stability of the injected planets. To rule out a potential planet, one of the two following conditions must be met: (i) if it exists, the planet signal should have been confidently detected in the RV data already; or (ii) its orbit would be dynamically non-viable. In other words, below the data-driven detection limits, any unstable area of the period-mass space should be further excluded and the detection limits would then  be adapted. While orbital stability is a well-established constraint to refine the orbital solutions of planetary systems \citep[e.g.][]{Tamayo2021, Stalport2022a}, ARDENT aims to export this strategy to the computation of detection limits. 

The detectability of a planet in the data is supposed independent from its dynamical plausibility.  Therefore, the probability of missing an additional planet can be expressed as 
\begin{equation} \label{eq:DLProbability}
\mathcal{P}_{miss} ~ = ~ (1-\mathcal{P}_{d}) \times \mathcal{P}_{s}, 
\end{equation}
where $\mathcal{P}_d$ is the probability of detecting the planet in the data and $\mathcal{P}_s$ is the probability that the resulting planetary system is stable. No matter the detectability of a hypothetical planet, it cannot exist if the resulting system is fully unstable ($\mathcal{P}_s=0$ meaning $\mathcal{P}_{miss}=0$). Conversely, if $\mathcal{P}_{s}$>0, the injected planet cannot be missed only if its detection rate is maximal ($\mathcal{P}_{d}$=1 meaning $\mathcal{P}_{miss}=0$). The aim of ARDENT is to compute the mass detection limits under the conditions of detectability and dynamical plausibility. 

\begin{figure}
    \centering
\includegraphics[width=\columnwidth]{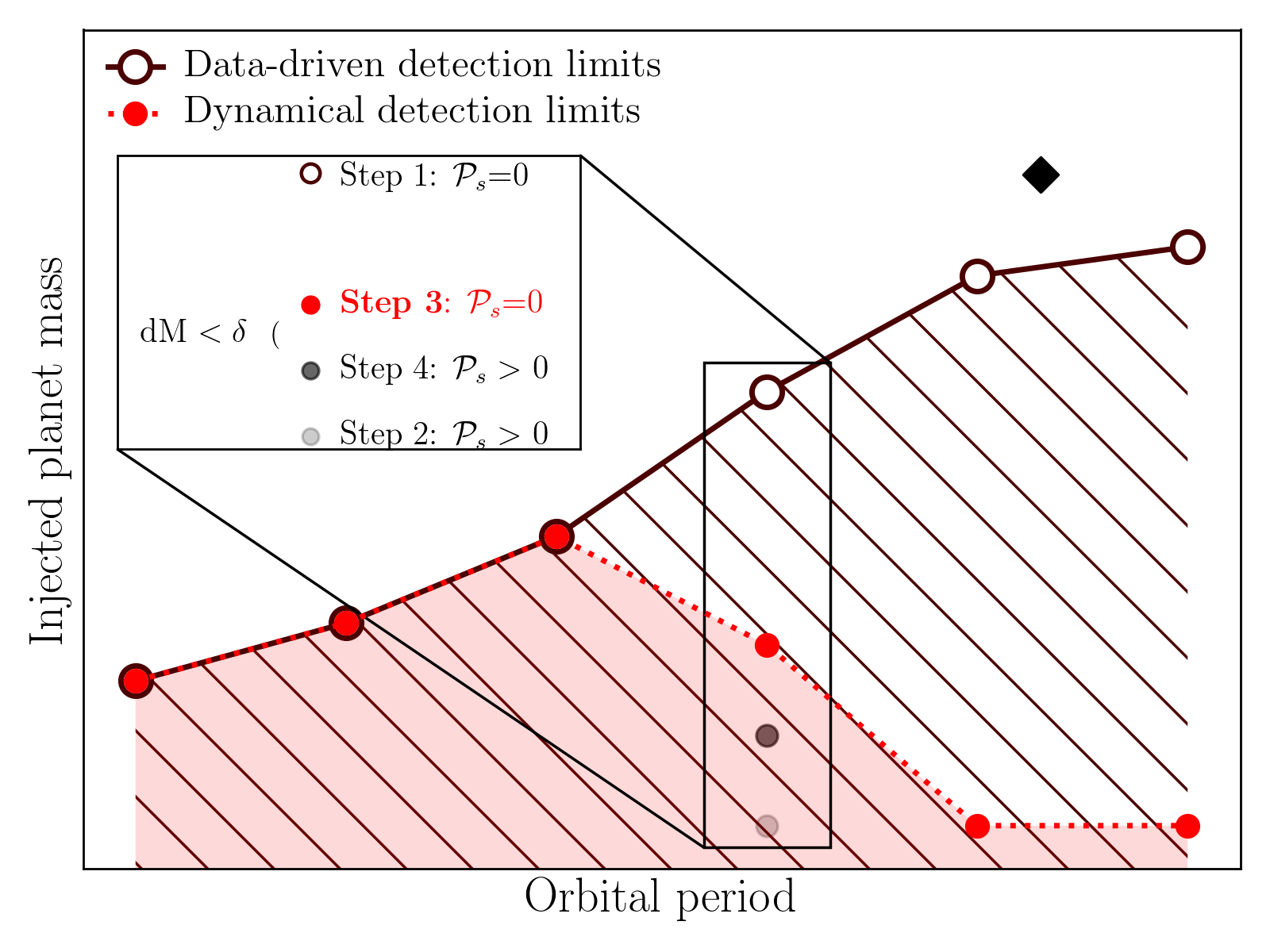}
\caption{Schematics illustrating dynamical detection limits computation in ARDENT. Empty circles represent the data-driven detection limits, above which $\mathcal{P}_{d}\sim1$. ARDENT models the dynamical influence of known planets in the system (diamond symbol), inducing an instability zone where additional planets are excluded. The resulting dynamical detection limits are shown in red. The inner panel illustrates the iterative mass limits update accounting for stability, with $dM$ the difference in injected mass between two consecutive steps and $\delta$ the precision threshold (cf. Sect. \ref{Sect:MassSampling}).}
\label{Fig:Schematics_DetectionLimits} 
\end{figure}

The computation of the data-driven detection limits in ARDENT is described in Sect. \ref{sect:data-drivenLimits}. By default, ARDENT computes the 95$\%$ detection limits. Above this limit curve, $\mathcal{P}_{miss}\sim0$ no matter the dynamical plausibility, $\mathcal{P}_{s}$, of the injected planets. Below this curve, $\mathcal{P}_{miss}\sim0$ exclusively when $\mathcal{P}_{s}\sim0$. Consequently, ARDENT only investigates the stability constraint below the data-driven detection limits. In the period-mass space where the detection limits are computed, the orbital stability depends on both parameters. For a given orbital period, the more massive the injected planet, the stronger the planet-planet gravitational interactions that can ultimately lead to orbital instability. 
Therefore, ARDENT searches for the minimal mass (below the data-driven limit curve) above which the resulting system is unstable. This limit mass defines the dynamical detection limits and pushes down the system completeness to lower masses. Figure \ref{Fig:Schematics_DetectionLimits} illustrates this general strategy. 
We cannot further rule out hidden planet candidates below the data-driven limit curve when they are not fully unstable, namely, when $\mathcal{P}_{s}$>0.

\subsection{Data-driven detection limits} \label{sect:data-drivenLimits}
The data-driven detection limits are the most common investigation in completeness studies. To compute them, ARDENT performs injection-recovery tests in the residual RV time series, namely, the time series that is left after subtracting the best Keplerian+activity model and that serves as input to the code. Planets are injected randomly in a range of periods, $P$, and RV semi-amplitudes, $K$. 
The range of $K$ is expressed in units of root-mean-square (rms) to decrease the instrumental dependence and provide a default range valid for most applications. Furthermore, a uniform sampling of $K$ results in a more efficient exploration of the period-mass space compared to a direct mass sampling (cf. Appendix \ref{App:Ksampling}). 

For each sampled $P$ and $K$, ARDENT injects the planet with a set of $N_p$ different orbital phase values, thereby performing $N_p$ injection-recovery tests for every point in the $P$-$K$ space. ARDENT injects each signal in the time series and attempts at recovering it via the computation of the generalised Lomb-Scargle (GLS) periodogram. The injected planet is considered as recovered if the most significant signal in the GLS periodogram has a false alarm probability (FAP) below 1$\%$ with a period estimation less than 5$\%$ away from the injected period. The detection rate of the sampled planet is thus given by its number of detections over the $N_p$ tests. Repeating this process on the entire sample of $P$ and $K$ fills up the parameter space within which the detection limits are computed. 

At the end of the injection-recovery tests, the period, mass (or $K$), and detection rate are used to compute the detection limits. ARDENT splits the sample into period bins. Given the period was sampled uniformly in log scale, each bin should contain approximately the same number of points if the number of injection-recovery tests is large enough. For each period bin, ARDENT fits for the cumulative distribution function (cdf) of the detection rate as a function of the injected mass. The fitted cdf is used to estimate the limit mass for a given detection threshold. By default, ARDENT returns the 95$\%$ detection limit. Repeating this process on each period bin leads to a detection limit curve in the period-mass space. These are the data-driven detection limits returned by ARDENT, which are used to investigate the stability constraint. 

\subsection{Dynamical detection limits} \label{sect:dynamicalLimits} 
The dynamical detection limits add the stability constraint to the condition of detectability of a hypothetical planet. In this framework, the probability of missing a planet is null if either its detectability is granted or if it is dynamically not viable (cf. Eq. \ref{eq:DLProbability}). This second condition allows to exclude zones of the parameter space where additional undetected planets would be unstable. As such, exclusively below the data-driven detection limits can the dynamical constraints make a difference, where $\mathcal{P}_{miss}=0$ only if $\mathcal{P}_{s}=0$. Investigating the orbital stability of a planetary system can be computationally expensive. The motivation for a fast inclusion of the stability constraint is what shapes the process of dynamical detection limits in ARDENT.

\subsubsection{Selective mass and period samplings} \label{Sect:MassSampling}
For a given set of orbital parameters, increasing the mass of an injected planet increases also the planet-planet gravitational interactions and the risk of orbital instabilities. Therefore, ARDENT searches for new mass limits below the data-driven curve that transition from stability to instability. 

To proceed, a selective mass sampling is applied to each period bin. The code first tests the stability of an injected planet with a mass of the data-driven detection limit (cf. Sect. \ref{Sect:StabCriteria}) and for an even sample of orbital phases. If it is stable for at least one phase angle, that point of the period-mass space cannot be excluded and the dynamical constraint does not provide further information than the RV data. Conversely, if the injected planet is unstable for all phase angles, it can be rejected and an updated mass limit has to be found. The code then tests the stability of a negligible mass (cf. Fig. \ref{Fig:Schematics_DetectionLimits}, step 2). Either the result remains unstable ($\mathcal{P}_{s}=0$), meaning no planet can exist at the period tested and the updated mass limit drops to zero; or the result is stable. In this case, ARDENT initiates a convergence process to find the mass transitioning from instability (step 1) to stability (step 2). At each new step, it injects planets with a mass in-between the nearest stable-unstable pair until a precision threshold on the mass, $\delta$, is reached. By default, we set $\delta$=0.5M$_{\oplus}$, which is lower than the precision of most RV instruments. This process is repeated independently for each period bin. Hence, it can be easily multi-processed. A built-in ARDENT parameter defines the number of local cores desired to compute the dynamical constraints. 

Regarding the scan in orbital periods, ARDENT employs an adaptive sampling. An initial loose-period binning (15 bins by default, equally spaced in log-scale) was used to compute the data-driven mass limits from the injection-recovery tests performed prior. Additionally, the sampling is significantly increased around each known planet. Specifically, ARDENT generates 100 more bins between $0.5P$ and $2P$, then interpolates the data-driven mass limits at each new bin. Such dense sampling around the known planets aims to resolve the main first-order mean-motion resonances (MMRs), which are particularly relevant in compact systems of small planets. ARDENT computes the updated mass limit for each period bin.

\subsubsection{Analytical and numerical stability criteria} \label{Sect:StabCriteria}
For each stability test, analytical criteria were first employed to identify unambiguously stable or unstable systems. To proceed, the angular momentum deficit (AMD) was computed \citep{Laskar2017}. It measures the difference between the angular momentum of the system and its equivalent with circular and co-planar orbits only. Within this framework, it is possible to define a threshold AMD below which even the worst redistribution of angular momentum among the planets would not lead to close encounters, ensuring the Hill stability \citep{Petit2018}. Alternatively, the AMD framework can be used to identify the region of the parameter space where MMRs overlap, leading to strong instability \citep{Petit2017}. Furthermore, a system is also classified as unstable if the initial planetary orbits cross but are not close to a MMR. We provide a detailed description of these analytical tests in Appendix \ref{App:AnalyticalCriteria}.  

When a system's stability cannot be clearly classified, further numerical investigations are carried out. ARDENT employs the REBOUND Python package \citep{Rein2012} with the \texttt{WHFast} symplectic integrator \citep{Rein2015} to perform numerical integrations over $10^4$ orbits of the outermost planet. If needed, a general relativity (GR) correction can be activated via REBOUNDx \citep{Tamayo2020} and the \texttt{gr} force structure. If the system survived the simulation (i.e. no close encounter, nor escape), the semi-major axes drift is computed based on 2000 evenly spaced outputs, from which the stability of the system is estimated. We provide further details and a calibration of stability on this drift in Appendix \ref{App:CalibDriftThr}.

\section{Application to TOI-1736} \label{Sect:TOI-1736} 
TOI-1736 is a solar analogue around which a transiting sub-Neptune and an outer giant planet have been found \citep{Martioli2023}. Investigating the multiplicity of inner SP systems in the presence of CJs is very informative. If the latter prevented the migration of SPs or their building blocks, a CJ-multiplicity correlation would be expected \citep[e.g.][]{Izidoro2015, Bonomo2025}. This trend might be even stronger in eccentric CJ systems, where dynamically active past evolution with planet-planet scattering likely occurred \citep{Huang2017}. Refining the completeness of systems such as TOI-1736 is of prime importance in tackling these open questions.  

\citet{Martioli2023} used 152 RV observations taken with the SOPHIE spectrograph to measure the planet masses. TOI-1736 b orbits the host star in 7.07 days with a mass and radius of 12.8$\pm$1.8 $M_{\oplus}$ and 2.44$\pm$0.18 $R_{\oplus}$. TOI-1736 c has a period and mass of 570.1$\pm$0.7 days and 8.7$\substack{+1.5 \\ -0.6}$ $M_{Jup}$, and an orbital inclination $i_c$=$69\substack{+15 \\ -17}$ deg measured with \textit{Gaia} DR3 excess noise \citep{Lindegren2021}. 
The orbital eccentricity is well constrained on TOI-1736 c ($e_c$=0.362$\pm$0.003), but it is not significant on planet b \citep[$e_b$ < 0.21, ][]{Martioli2023}. The over-estimation of eccentricities is a common issue in RV observations, particularly for small-amplitude signals \citep{Hara2019}. In the context of ARDENT, this erroneously biases the detection limits by enlarging the instability zones around the planets. To prevent this issue, the parameter \texttt{elower} indicates the lower limit of the eccentricity compatible with the observations. When specified (recommended when uncertainties are significant), ARDENT automatically switches to this value to model the system dynamics. As such, the eccentricity of TOI-1736 b is considered to be null.

\begin{figure}
    \centering
\includegraphics[width=\columnwidth]{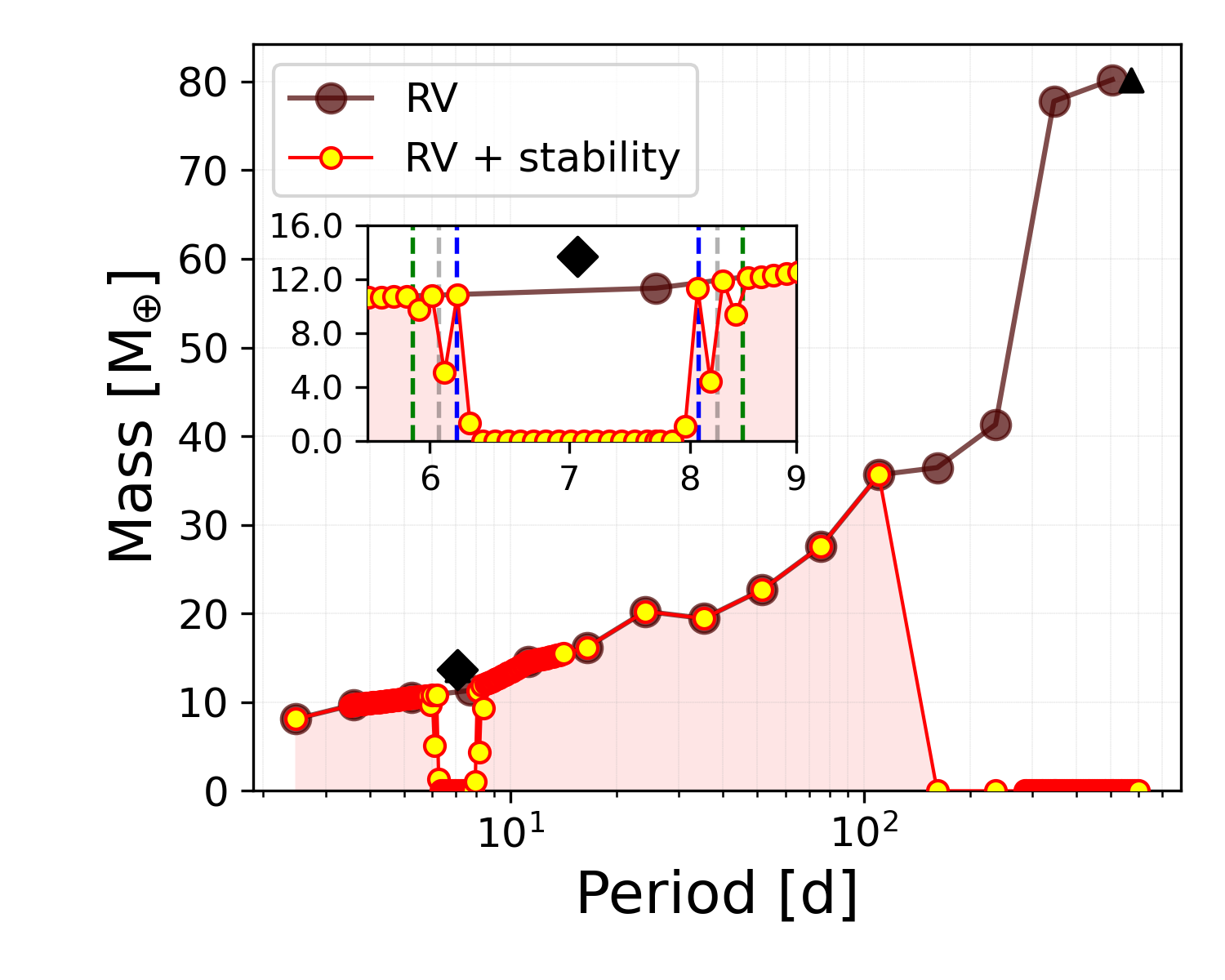}
\caption{Data-driven (in brown) and dynamical (in yellow-red) detection limits of TOI-1736 in the period-mass space, derived from SOPHIE RVs. Known planets are shown in black (triangle symbol if the mass is beyond the plotted range). Additional planets may only exist in the red-shaded area. A zoomed-in window around the inner planet illustrates nearby low-order MMRs (green: 6:5 ; gray: 7:6 ; blue: 8:7).} 
\label{Fig:TOI1736_DLcurve} 
\end{figure}

We retrieved the public SOPHIE RV data, and fit the time series with a model including two Keplerians and a linear drift\footnote{To fit the data, we used the Data $\&$ Analysis Center for Exoplanets (DACE, \url{https://dace.unige.ch}).}, similarly  to the approach presented in \citet{Martioli2023}. Injection-recovery tests were performed on the residual RVs to compute the data-driven mass limits in the period range [2, 600] days. In total, 2000 samples in the $P$-$K$ space were generated, each composed of ten tests with different orbital phases with an even spread in $[-\pi,\pi[$. We retrieved the 95$\%$ data-driven detection limits from these tests, which were then used together with the known planet parameters to compute the dynamical detection limits. Using the default ARDENT settings and four cores on a MacBook Pro equipped with the M2 chip, the dynamical limits computation took about 20 minutes. The result is presented in Fig. \ref{Fig:TOI1736_DLcurve}. Appendix \ref{App:CodeTOI1736} details the command lines. 

Strikingly, the outer CJ prevents any additional inner planet to exist beyond an orbital period of $\sim$150 days. This provides an upper limit to the inner system. Additionally, a small instability zone is observed in the neighbourhood of TOI-1736 b, also impacting  the detection limits. The effect of nearby MMR overlap is visible, as highlighted by the vertical dashed lines. Overall, the inclusion of the stability constraint significantly strengthens the completeness of the system. Furthermore, this result is of high importance for RV follow-up programs. Indeed, by reducing the period window of potential additional planets, the observational strategy can be adapted consequently. An alternative view of these detection limits, in the orbital plane, is presented in Appendix \ref{App:CodeTOI1736}. Finally, we also applied ARDENT in the period range between 100 and 3000 days. While data-driven mass limits are saturated at periods larger than the RV baseline ($\sim$ 900 days), the stability constraint informs the range of forbidden periods around the giant TOI-1736 c. We find that no additional planet below 1500 days can exist in the system.

\section{Conclusions} \label{Sect:Conclusion}
The impact of CJs on SP formation and evolution remains an open question, placing the Solar System in the exoplanet context. Soon, data from \textit{Gaia} DR4, PLATO, and subsequent RV follow-ups are likely to dramatically increase the number of SP-CJ systems. In this context, benefiting from the strongest systems completeness with the data at hand will be key to reveal CJ-SP (anti-)correlations. It is therefore timely to have a tool that computes the detection limits with the stability constraints. In this Letter, we introduce ARDENT, a publicly available Python package that derives both data-driven and dynamical detection limits. ARDENT is designed to be fast, thanks to a selective exploration of the period-mass space and the combination of analytical and numerical stability criteria. To test the method, we analysed TOI-1736, a system containing a 7-day sub-Neptune and a 570-day eccentric cold giant, from the publicly available SOPHIE RVs. This study highlighted the advantages of ARDENT. Within a limited amount of computation time, we were able to strengthen the system completeness and provide RV follow-up programs with a reduced period window for potential additional planets. The perspective of employing ARDENT on a sample of CJ systems is expected to improve the precision in SP occurrence rates, a crucial step to revealing SP-CJ correlations. While CJ systems are well-suited for ARDENT, the code can be used on a diverse range of system architectures, including compact systems or eccentric planets. The package is user-friendly, and can be automatised to derive detection limits on a large sample of systems\footnote{Further information on installation and tutorials can be found at \url{https://github.com/manustalport/ardent}.}.

\begin{acknowledgements}  
We thank the referee for insightful feedback on the code and manuscript. M.S. acknowledges financial support from the Belgian Federal Science Policy Office (BELSPO) in the framework of the PRODEX Programme of the European Space Agency (ESA) under contract number C4000140754. 
L.N. acknowledges financial contribution from the INAF Large Grant 2023 "EXODEMO".
V.V.G. is an F.R.S.-FNRS Research Associate. 
\end{acknowledgements}

\bibliographystyle{aa} 
\bibliography{bib.bib}

\begin{thebibliography}{25}
\expandafter\ifx\csname natexlab\endcsname\relax\def\natexlab#1{#1}\fi

\bibitem[{{Barbato} {et~al.}(2023){Barbato}, {S{\'e}gransan}, {Udry}, {Unger},
  {Bouchy}, {Lovis}, {Mayor}, {Pepe}, {Queloz}, {Santos}, {Delisle},
  {Figueira}, {Marmier}, {Matthews}, {Lo Curto}, {Venturini}, {Chaverot},
  {Cretignier}, {Otegi}, \& {Stalport}}]{Barbato2023}
{Barbato}, D., {S{\'e}gransan}, D., {Udry}, S., {et~al.} 2023, \aap, 674, A114

\bibitem[{{Barbato} {et~al.}(2018){Barbato}, {Sozzetti}, {Desidera}, {Damasso},
  {Bonomo}, {Giacobbe}, {Colombo}, {Lazzoni}, {Claudi}, {Gratton}, {LoCurto},
  {Marzari}, \& {Mordasini}}]{Barbato2018}
{Barbato}, D., {Sozzetti}, A., {Desidera}, S., {et~al.} 2018, \aap, 615, A175

\bibitem[{{Bonomo} {et~al.}(2023){Bonomo}, {Dumusque}, {Massa}, {Mortier},
  {Bongiolatti}, {Malavolta}, {Sozzetti}, {Buchhave}, {Damasso}, {Haywood},
  {Morbidelli}, {Latham}, {Molinari}, {Pepe}, {Poretti}, {Udry}, {Affer},
  {Boschin}, {Charbonneau}, {Cosentino}, {Cretignier}, {Ghedina}, {Lega},
  {L{\'o}pez-Morales}, {Margini}, {Mart{\'\i}nez Fiorenzano}, {Mayor},
  {Micela}, {Pedani}, {Pinamonti}, {Rice}, {Sasselov}, {Tronsgaard}, \&
  {Vanderburg}}]{Bonomo2023}
{Bonomo}, A.~S., {Dumusque}, X., {Massa}, A., {et~al.} 2023, \aap, 677, A33

\bibitem[{{Bonomo} {et~al.}(2017){Bonomo}, {H{\'e}brard}, {Raymond}, {Bouchy},
  {Lecavelier des Etangs}, {Bord{\'e}}, {Aigrain}, {Almenara}, {Alonso},
  {Cabrera}, {Csizmadia}, {Damiani}, {Deeg}, {Deleuil}, {D{\'\i}az}, {Erikson},
  {Fridlund}, {Gandolfi}, {Guenther}, {Guillot}, {Hatzes}, {Izidoro}, {Lovis},
  {Moutou}, {Ollivier}, {P{\"a}tzold}, {Rauer}, {Rouan}, {Santerne}, \&
  {Schneider}}]{Bonomo2017}
{Bonomo}, A.~S., {H{\'e}brard}, G., {Raymond}, S.~N., {et~al.} 2017, \aap, 603,
  A43

\bibitem[{{Bonomo} {et~al.}(2025){Bonomo}, {Naponiello}, {Pezzetta},
  {Sozzetti}, {Gandolfi}, {Wittenmyer}, \& {Pinamonti}}]{Bonomo2025}
{Bonomo}, A.~S., {Naponiello}, L., {Pezzetta}, E., {et~al.} 2025, \aap, 700,
  A126

\bibitem[{{Brewer} {et~al.}(2020){Brewer}, {Fischer}, {Blackman}, {Cabot},
  {Davis}, {Laughlin}, {Leet}, {Ong}, {Petersburg}, {Szymkowiak}, {Zhao},
  {Henry}, \& {Llama}}]{Brewer2020}
{Brewer}, J.~M., {Fischer}, D.~A., {Blackman}, R.~T., {et~al.} 2020, \aj, 160,
  67

\bibitem[{{Bryan} \& {Lee}(2024)}]{Bryan2024}
{Bryan}, M.~L. \& {Lee}, E.~J. 2024, \apjl, 968, L25

\bibitem[{{Dai} {et~al.}(2024){Dai}, {Goldberg}, {Batygin}, {van Saders},
  {Chiang}, {Choksi}, {Li}, {Petigura}, {Gilbert}, {Millholland}, {Dai},
  {Bouma}, {Weiss}, \& {Winn}}]{Dai2024}
{Dai}, F., {Goldberg}, M., {Batygin}, K., {et~al.} 2024, \aj, 168, 239

\bibitem[{{Hara} {et~al.}(2019){Hara}, {Bou{\'e}}, {Laskar}, {Delisle}, \&
  {Unger}}]{Hara2019}
{Hara}, N.~C., {Bou{\'e}}, G., {Laskar}, J., {Delisle}, J.~B., \& {Unger}, N.
  2019, \mnras, 489, 738

\bibitem[{{Huang} {et~al.}(2017){Huang}, {Petrovich}, \& {Deibert}}]{Huang2017}
{Huang}, C.~X., {Petrovich}, C., \& {Deibert}, E. 2017, \aj, 153, 210

\bibitem[{{Izidoro} {et~al.}(2015){Izidoro}, {Raymond}, {Morbidelli},
  {Hersant}, \& {Pierens}}]{Izidoro2015}
{Izidoro}, A., {Raymond}, S.~N., {Morbidelli}, A., {Hersant}, F., \& {Pierens},
  A. 2015, \apjl, 800, L22

\bibitem[{{Kopparapu} {et~al.}(2013){Kopparapu}, {Ramirez}, {Kasting}, {Eymet},
  {Robinson}, {Mahadevan}, {Terrien}, {Domagal-Goldman}, {Meadows}, \&
  {Deshpande}}]{Kopparapu2013}
{Kopparapu}, R.~K., {Ramirez}, R., {Kasting}, J.~F., {et~al.} 2013, \apj, 765,
  131

\bibitem[{{Laskar} \& {Petit}(2017)}]{Laskar2017}
{Laskar}, J. \& {Petit}, A.~C. 2017, \aap, 605, A72

\bibitem[{{Lindegren} {et~al.}(2021){Lindegren}, {Klioner}, {Hern{\'a}ndez},
  {Bombrun}, {Ramos-Lerate}, {Steidelm{\"u}ller}, {Bastian}, {Biermann}, {de
  Torres}, {Gerlach}, {Geyer}, {Hilger}, {Hobbs}, {Lammers}, {McMillan},
  {Stephenson}, {Casta{\~n}eda}, {Davidson}, {Fabricius}, {Gracia-Abril},
  {Portell}, {Rowell}, {Teyssier}, {Torra}, {Bartolom{\'e}}, {Clotet},
  {Garralda}, {Gonz{\'a}lez-Vidal}, {Torra}, {Abbas}, {Altmann}, {Anglada
  Varela}, {Balaguer-N{\'u}{\~n}ez}, {Balog}, {Barache}, {Becciani}, {Bernet},
  {Bertone}, {Bianchi}, {Bouquillon}, {Brown}, {Bucciarelli}, {Busonero},
  {Butkevich}, {Buzzi}, {Cancelliere}, {Carlucci}, {Charlot}, {Cioni},
  {Crosta}, {Crowley}, {del Peloso}, {del Pozo}, {Drimmel}, {Esquej}, {Fienga},
  {Fraile}, {Gai}, {Garcia-Reinaldos}, {Guerra}, {Hambly}, {Hauser},
  {Jan{\ss}en}, {Jordan}, {Kostrzewa-Rutkowska}, {Lattanzi}, {Liao}, {Licata},
  {Lister}, {L{\"o}ffler}, {Marchant}, {Masip}, {Mignard}, {Mints}, {Molina},
  {Mora}, {Morbidelli}, {Murphy}, {Pagani}, {Panuzzo}, {Pe{\~n}alosa Esteller},
  {Poggio}, {Re Fiorentin}, {Riva}, {Sagrist{\`a} Sell{\'e}s}, {Sanchez
  Gimenez}, {Sarasso}, {Sciacca}, {Siddiqui}, {Smart}, {Souami}, {Spagna},
  {Steele}, {Taris}, {Utrilla}, {van Reeven}, \& {Vecchiato}}]{Lindegren2021}
{Lindegren}, L., {Klioner}, S.~A., {Hern{\'a}ndez}, J., {et~al.} 2021, \aap,
  649, A2

\bibitem[{{Martioli} {et~al.}(2023){Martioli}, {H{\'e}brard}, {de Almeida},
  {Heidari}, {Lorenzo-Oliveira}, {Kiefer}, {Almenara}, {Bieryla}, {Boisse},
  {Bonfils}, {Brice{\~n}o}, {Collins}, {Cort{\'e}s-Zuleta}, {Dalal}, {Deleuil},
  {Delfosse}, {Demangeon}, {Eastman}, {Forveille}, {Furlan}, {Howell}, {Hoyer},
  {Jenkins}, {Latham}, {Law}, {Mann}, {Moutou}, {Santos}, {Sousa}, {Stassun},
  {Stockdale}, {Torres}, {Twicken}, {Winn}, \& {Ziegler}}]{Martioli2023}
{Martioli}, E., {H{\'e}brard}, G., {de Almeida}, L., {et~al.} 2023, \aap, 680,
  A84

\bibitem[{{Petit} {et~al.}(2017){Petit}, {Laskar}, \& {Bou{\'e}}}]{Petit2017}
{Petit}, A.~C., {Laskar}, J., \& {Bou{\'e}}, G. 2017, \aap, 607, A35

\bibitem[{{Petit} {et~al.}(2018){Petit}, {Laskar}, \& {Bou{\'e}}}]{Petit2018}
{Petit}, A.~C., {Laskar}, J., \& {Bou{\'e}}, G. 2018, \aap, 617, A93

\bibitem[{{Rein} \& {Liu}(2012)}]{Rein2012}
{Rein}, H. \& {Liu}, S.~F. 2012, \aap, 537, A128

\bibitem[{{Rein} \& {Tamayo}(2015)}]{Rein2015}
{Rein}, H. \& {Tamayo}, D. 2015, \mnras, 452, 376

\bibitem[{{Rosenthal} {et~al.}(2021){Rosenthal}, {Fulton}, {Hirsch},
  {Isaacson}, {Howard}, {Dedrick}, {Sherstyuk}, {Blunt}, {Petigura}, {Knutson},
  {Behmard}, {Chontos}, {Crepp}, {Crossfield}, {Dalba}, {Fischer}, {Henry},
  {Kane}, {Kosiarek}, {Marcy}, {Rubenzahl}, {Weiss}, \&
  {Wright}}]{Rosenthal2021}
{Rosenthal}, L.~J., {Fulton}, B.~J., {Hirsch}, L.~A., {et~al.} 2021, \apjs,
  255, 8

\bibitem[{{Stalport} {et~al.}(2022){Stalport}, {Delisle}, {Udry}, {Matthews},
  {Bourrier}, \& {Leleu}}]{Stalport2022a}
{Stalport}, M., {Delisle}, J.~B., {Udry}, S., {et~al.} 2022, \aap, 664, A53

\bibitem[{{Tamayo} {et~al.}(2021){Tamayo}, {Gilbertson}, \&
  {Foreman-Mackey}}]{Tamayo2021}
{Tamayo}, D., {Gilbertson}, C., \& {Foreman-Mackey}, D. 2021, \mnras, 501, 4798

\bibitem[{{Tamayo} {et~al.}(2020){Tamayo}, {Rein}, {Shi}, \&
  {Hernandez}}]{Tamayo2020}
{Tamayo}, D., {Rein}, H., {Shi}, P., \& {Hernandez}, D.~M. 2020, \mnras, 491,
  2885

\bibitem[{{Tamayo} {et~al.}(2016){Tamayo}, {Silburt}, {Valencia}, {Menou},
  {Ali-Dib}, {Petrovich}, {Huang}, {Rein}, {van Laerhoven}, {Paradise},
  {Obertas}, \& {Murray}}]{Tamayo2016}
{Tamayo}, D., {Silburt}, A., {Valencia}, D., {et~al.} 2016, \apjl, 832, L22

\bibitem[{{Zhu} {et~al.}(2018){Zhu}, {Petrovich}, {Wu}, {Dong}, \&
  {Xie}}]{Zhu2018}
{Zhu}, W., {Petrovich}, C., {Wu}, Y., {Dong}, S., \& {Xie}, J. 2018, \apj, 860,
  101

\end{thebibliography}

\begin{appendix} 
\section{Sampling planets for the injection-recovery tests} \label{App:Ksampling}
To compute the data-driven detection limits, ARDENT proceeds to injection-recovery tests in the period versus RV semi-amplitude space. It generates a sample of injected planets by drawing $P$ and $K$ from log-uniform and uniform distributions, respectively. The range of $K$ is expressed in units of rms (by default, [0.1; 1.2] rms inside which $K$ is drawn uniformly). As such, it does not need adjustments from an instrument to another, but the user can modify this input range if desired. 

\begin{figure}[!h]
    \centering
\includegraphics[width=\columnwidth]{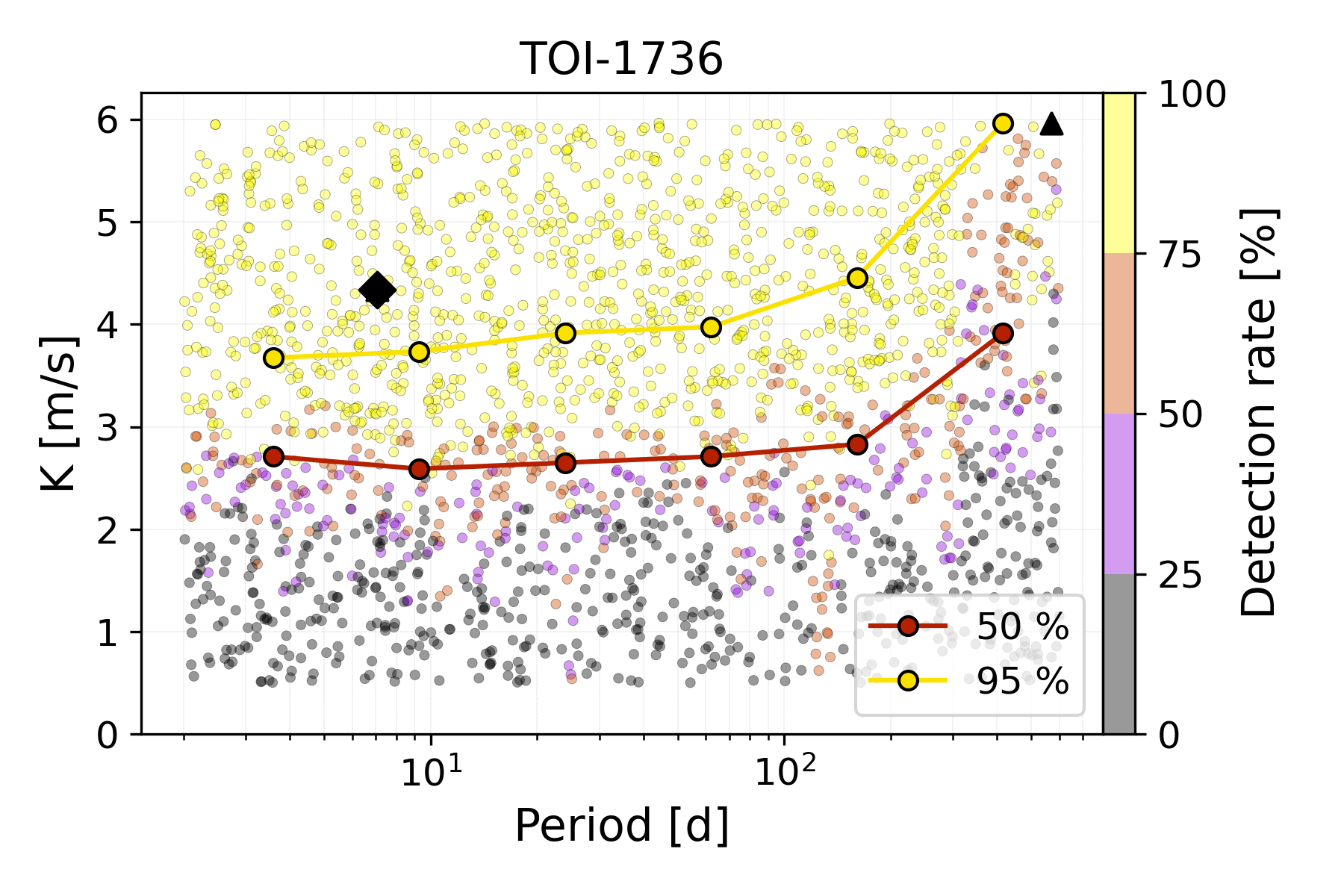}
\caption{TOI-1736 injection-recovery tests in the period versus $K$ space. The detection limits (here the 50 and 95$\%$ limits are represented) are constant in $K$, until the RV timespan becomes too short compared to the injected periods. Consequently, a uniform sampling in $K$ is efficient to sample the zone close to the detection limits.} 
\label{Fig:TOI1736_P-K} 
\end{figure}

Interestingly, the data-driven detection limits in $K$ do not depend on the orbital period, but only on the timespan of the observations. In other words, the detection limit along $K$ is a constant. This is illustrated in Fig. \ref{Fig:TOI1736_P-K}, which presents the ARDENT output of the data-driven $K$ detection limits for TOI-1736 analysed in this work. Each point in this period-$K$ space corresponds to a set of ten different injected planets, each of them with a different orbital phase. The colour code depicts the detection rate as the number of detected planets over the ten different tests per ($P$, $K$). The detection limits stand at a constant value of $K$, until the RV timespan becomes an intrinsic limitation for the detectability. 

\begin{figure}
    \centering
\includegraphics[width=\columnwidth]{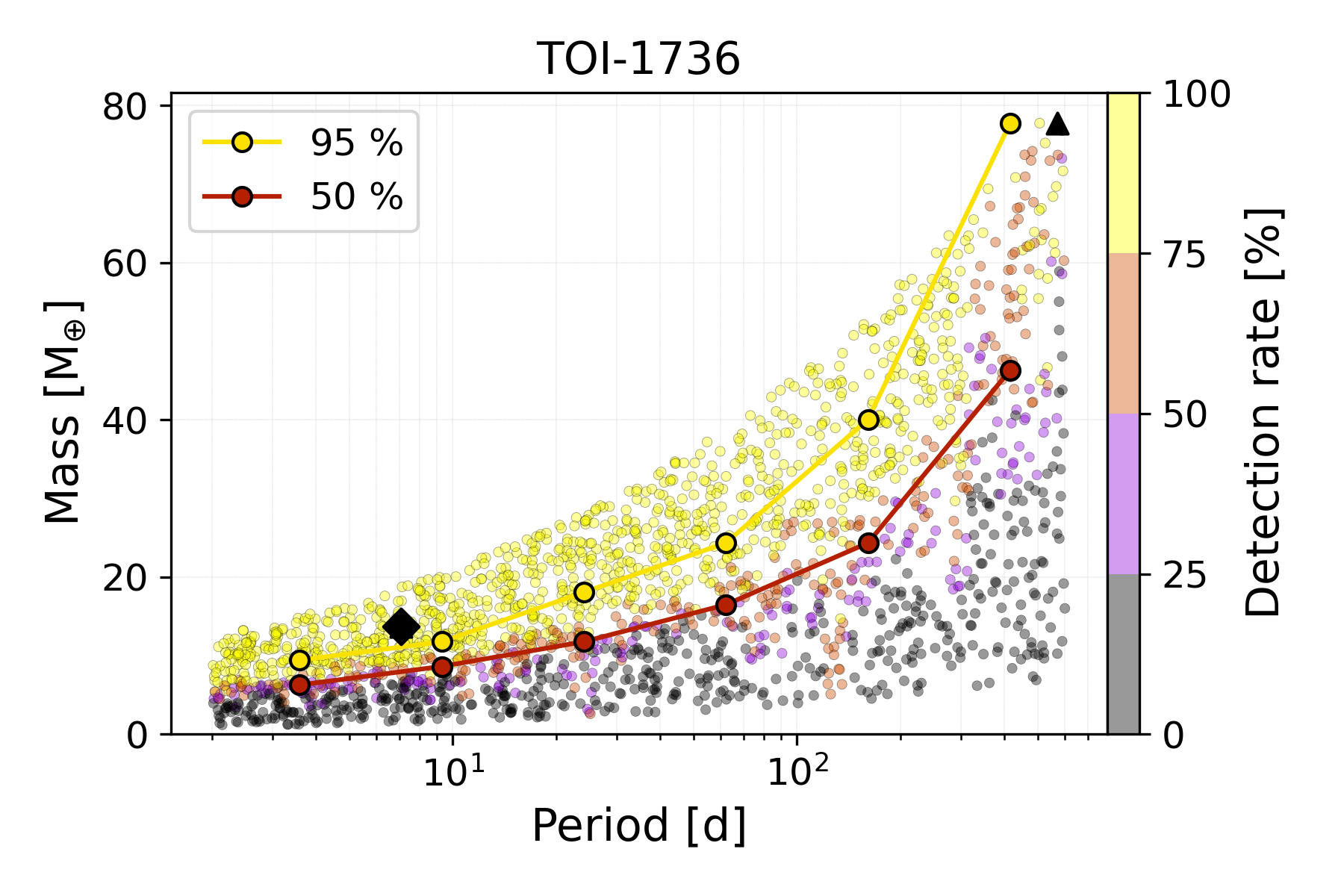}
\caption{TOI-1736 injection-recovery tests in the period-mass space, obtained after converting the $K$ distribution (cf. Fig. \ref{Fig:TOI1736_P-K}) into mass. Because of the efficient sampling in $K$, this mass distribution is also efficient to sample the detection limits.} 
\label{Fig:TOI1736_P-M} 
\end{figure}

Since detection limits are constant in the ($P$,$K$) space (considering long enough RV time spans), a uniform sampling in $K$ is particularly efficient to sample the injected planet mass. Indeed, it insures that the sampled mass is always close to the desired detection limit. After converting the $K$ distribution into planet mass, we obtain the plot shown in Fig. \ref{Fig:TOI1736_P-M} (which is also a direct output of ARDENT). This distribution of planets in the period-mass space samples densely the useful region surrounding the detection limits, as a direct consequence of the sampling in $K$. 
From this period-mass sample, the data-driven mass limits are computed by fitting the cumulative distribution function (cdf) of the detection rate as a function of injected mass, for each period bin. Because this sampling is dense around the transition between 0 and 100$\%$ detection rates, the cdf is also well sampled. 

\FloatBarrier

\section{Analytical stability criteria} \label{App:AnalyticalCriteria}
ARDENT employs analytical criteria to estimate the orbital stability of planetary systems. These enable us to identify long-term stable and short-term unstable systems. If a clear answer about stability cannot be provided, the code initiates numerical investigations. In this section we expand on the description of the analytical tests. 

\subsection{Orbit crossings}
ARDENT checks for crossing orbits on all the planet pairs in the system via the simple expression  
\begin{equation}
a_{i} (1+e_i) < a_o (1-e_o), 
\end{equation} 
where the subscripts $i$ and $o$ refer to the inner and outer planets of the pair, and $a$ and $e$ are the initial semi-major axis and orbital eccentricity, respectively. When this condition is violated, the orbits cross. Unless the planet pair is in mean-motion resonance (MMR), the orbit crossing will ultimately lead to planet-planet close encounters and rapid instability. However, inside a MMR, the orbital phases can be protected from close encounters in crossing orbits (the example of the Neptune - Pluto pair in the Solar System). Therefore, crossing orbits are indicative of rapid instability only if the pair is not in resonance. 

To systematically check for MMR, we adopt the simple resonance criterion of \citet{Dai2024} based on the distance from the exact period commensurability. The latter is defined as  
\begin{equation}
\Delta \equiv \dfrac{P_{out}/P_{in}}{p/q} - 1, 
\end{equation}
where $P_{in}$ and $P_{out}$ are the orbital period of the inner and outer planet of the pair, and $p$ and $q$ are small integers such that $p-q$ gives the order of the resonance. Based on the \textit{Kepler} sample of multi-planet systems, \citet{Dai2024} define a distance threshold below which the planet pair is classified as resonant. In particular, for first-order MMRs ($q$=$p$-1), -0.015 < $\Delta$ < 0.03. For second and third-order MMRs ($q$=$p$-2 and $q$=$p$-3), -0.015 < $\Delta$ < 0.015. Within ARDENT, this criterion is used to check for the main low-order resonances: 
\begin{itemize}
    \item Order 1 MMRs: 2:1, 3:2, 4:3, 5:4, 6:5  
    \item Order 2 MMRs: 3:1, 5:3, 7:5 
    \item Order 3 MMRs: 4:1, 5:2  
\end{itemize}
As noted by \citet{Dai2024}, a significant amount of planet pairs satisfying this criterion do not have librating resonant arguments (yet, they likely underwent resonant capture in the past). This criterion is therefore optimistic in considering planet pairs in resonance. In the case of ARDENT, it implies that these close-to-MMR planet pairs will not be automatically rejected from the orbit crossing criterion, and will go through a numerical investigation. Very likely, orbital instability will then rapidly set in these non-resonant crossing systems. 

\subsection{AMD framework}
In case the system passed the orbits crossing test, another stability criterion is applied based on the AMD of the system \citep{Laskar2017}. The AMD measures the difference between the angular momentum of the system, and the one it would have if all the orbits were circular and co-planar. Hence it informs about the dynamical excitation of the system. The AMD of a system is given by summing the AMDs of its planets in the heliocentric coordinates \citep{Laskar2017}:  
\begin{equation}
AMD = \sum_{k=1}^{n} \Lambda_k \left(1 - \sqrt{1-e_k^2} \cos i_k \right), 
\end{equation}
where $\Lambda_k = M_k \sqrt{G M_{\star} a_k}$, $M_k$ and $M_{\star}$ are the planet and stellar masses, $G$ is the gravitational constant, $a_k$ is the semi-major axis of planet $k$, $e_k$, and $i_k$ are the orbital eccentricity and inclination, respectively. 

Within this framework, a system is long-term stable if any redistribution of the angular momentum among the planets cannot lead to close encounters, compatible with the Hill stability \citep{Petit2018}. Therefore, it is possible to define a critical AMD $C_c^H$ below which this condition is met and the system is Hill-AMD stable. This Hill-AMD criterion was investigated for each consecutive planet pair, and the system is AMD-stable if all the pairs are. Considering two planets with semi-major axes $a_1$ and $a_2$ such that $\alpha=\frac{a_1}{a_2}<1$, the critical Hill-AMD is given by \citep{Petit2018} 
\begin{equation}
C_c^H ~ = ~ \gamma\sqrt{\alpha} + 1 - (1+\gamma)^{3/2} \sqrt{\dfrac{\alpha}{\gamma+\alpha} \left(1+\dfrac{3^{4/3}\epsilon^{2/3}\gamma}{(1+\gamma)^2}\right)}, 
\end{equation}
where $\gamma=\frac{M_1}{M_2}$ and $\epsilon=\frac{M_1+M_2}{M_{\star}}$. The condition of Hill stability is verified if for every consecutive planet pair
\begin{equation}
\beta_H ~ \equiv ~ \dfrac{AMD / \Lambda_2}{C_c^H} ~ < ~ 1. 
\end{equation}

Alternatively, for close planet pairs, the resonance web becomes dense and a critical AMD $C_c^{MMR}$ based on the overlap of first-order MMRs can be defined \citep{Petit2017}. Above this limit, MMR overlaps induce rapid instability. The critical AMD in the context of MMR overlaps is given by \citep{Petit2017}
\begin{equation}
C_c^{MMR} ~ = ~ \dfrac{\gamma}{1+\gamma} \dfrac{g^2}{2}, 
\end{equation}
where $g = \frac{3^4(1-\alpha)^5}{2^9 r \epsilon} - \frac{32 r \epsilon}{9(1-\alpha)^2}$ and $r=0.80199$. The planet pair is AMD-unstable if 
\begin{equation}
\beta_{MMR} \equiv \dfrac{AMD / \Lambda_2}{C_c^{MMR}} > 1. 
\end{equation}
\citet{Petit2017} derived the threshold semi-major axes ratio $\alpha_R$ above which the MMR overlap criterion is to be considered. Below this threshold, the Hill criterion prevails. 

To synthesise, the AMD of consecutive planet pairs is computed and compared to $C_c^H$ or $C_c^{MMR}$ depending on the semi-major axes ratio $\alpha=a_1/a_2$ of the pair: 
\begin{itemize}
    \item $\alpha < \alpha_R$: stable system if $\beta_H$<1
    \item $\alpha > \alpha_R$: unstable system if $\beta_{MMR}$>1
\end{itemize}  

In conclusion, several analytical criteria are used in ARDENT to classify the stability of planetary systems. The systems are classified as 
\begin{enumerate}
    \item unstable if (i) orbits cross far from MMRs, or (ii) if the systems are located in the MMR overlap region ($\beta_{MMR}$>1);
    \item stable if they pass the Hill-AMD stability criterion ($\beta_H$<1); 
    \item undefined otherwise. In this case, a numerical investigation is initiated. 
\end{enumerate}  
Finally, it is worth noting that in systems containing an eccentric outer giant planet such as in TOI-1736, the majority of the injected planets are either unstable due to the orbits crossing criterion, or need further numerical investigation. Fewer injected planets are Hill-AMD stable, the reason being that any planet pair including the outer eccentric giant has a large AMD which penalises the Hill stability.

\FloatBarrier

\section{Calibration of the threshold in semi-major axis drift} \label{App:CalibDriftThr}
ARDENT classifies the stability of planetary systems based on analytic criteria at first, and via numerical simulations if needed. The latter consist of short integrations over $10^4$ orbits of the outermost planet, after which the level of chaos is estimated. Chaos is indicative of orbital instability \citep[e.g.][]{Stalport2022a} before that a close encounter or planet escape occurs. Therefore, it is an efficient way to speed up computation time. To estimate chaos in planetary systems, ARDENT measures the drift in semi-major axes. Under secular, non chaotic dynamics, the averaged semi-major axes are conserved quantities and do not vary on long time scales. In contrast, they drift significantly in chaotic dynamics. Therefore, the measure of the semi-major axes drift offers an estimation of chaos and, indirectly, orbital stability. As was shown by \citet{Tamayo2016}, the study of semi-major axes provides a reliable stability indicator. They trained a machine on several orbital features based on short ($10^4$ orbits) integrations to recognise patterns of instability, and compared those results with a control set derived from long-term ($10^7$ orbits) integrations. The variation in semi-major axes was ranked among the top features predicting (in)stability. 

In ARDENT, we define the relative drift in semi-major axis as
\begin{equation}
\Delta a ~ = ~ max_i \left(\dfrac{| a_{2,i}-a_{1,i} |}{a_{0,i}} \right), 
\end{equation}
with $a_{1,i}$ and $a_{2,i}$ as the median semi-major axis of planet $i$ over the first and second halves of the integration, respectively, and $a_{0,i}$ as the initial semi-major axis. The semi-major axis drift of the whole system is defined as the maximum relative drift amongst all planets. At the end of the numerical simulation, the planetary system is classified as unstable if either the integration did not reach the end (escape or close-encounter), or if one of the planetary orbits drifted by more than a certain threshold $\Delta a > \Delta a_{max}$. 

This threshold drift $\Delta a_{max}$ is an argument of the ARDENT's function computing the dynamical detection limits, and hence it can be easily adjusted. However, we carried out experiments to calibrate this threshold and motivate its default value in ARDENT. We generated three families of 1000 artificial 3-planet systems each, composed of two inner SP and an outer CJ. In all systems, the period and mass of the innermost and outermost planets are fixed at [10, 500]  days and [5 M$_{\oplus}$, 5 M$_{Jup}$], respectively. Regarding the middle planet, its period and mass are randomly drawn from log-uniform distributions in [11, 200] days and [0.1, 10] $\times M_b$, respectively. The inner SPs have null eccentricities. In the first set of systems, the CJ has a null eccentricity $e_{CJ}=0$ (the null eccentricity set). In the two other samples, $e_{CJ}=0.3$ and $0.7$ (the intermediate and high eccentricity sets, respectively). 

\begin{figure}
    \centering
\includegraphics[width=\columnwidth]{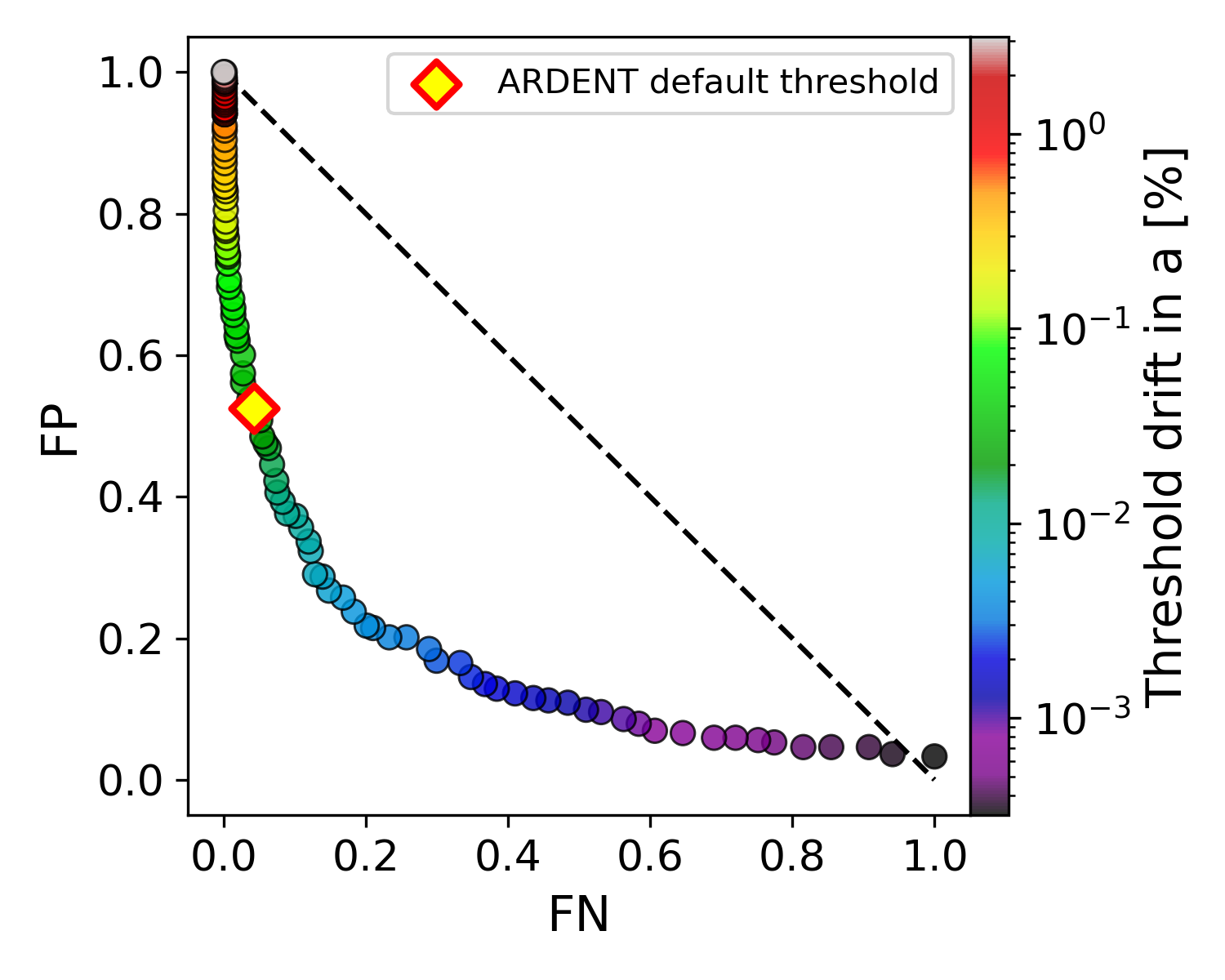}
\caption{Calibration of the numerical stability threshold employed in ARDENT. This plot results from short and long numerical integrations on three sets of systems, differing from each other with the eccentricity of the outer CJ. The relative amounts of FP and FN are shown for all sets combined, and for a various range of threshold drifts in semi-major axis $\Delta a_{max}$ indicated by the colorbar. The diamond symbol corresponds to the default ARDENT value of $\Delta a_{max}$=0.025$\%$. This conservative choice favours minimising the amount of FN.} 
\label{Fig:CalibDrift} 
\end{figure}

For every planetary system, we carried out two numerical simulations. We ran a short simulation over $10^4 \times P_{outer}$, similar to the ARDENT simulations time scale, after which $\Delta a$ was computed. Secondly, we ran a long simulation over 10$^7$ years for reference\footnote{We carried both simulations with REBOUND and its symplectic integrator \texttt{WHFast} with an integration time step of $P_{inner}/50$ and $P_{inner}/100$ for the short and long simulations, respectively.}. With an arbitrary drift threshold $\Delta a_{max}$, we could estimate each system stability from the short simulations and compare this classification to the real answer coming from the control sample of long integrations. Systems that were classified as unstable but survived the long integration were defined as false negatives (FN). Alternatively, systems wrongly classified as stable were defined as false positives (FP). We investigated the amounts of FN and FP for a various range of drift thresholds $\Delta a_{max}$, and present the result in Fig. \ref{Fig:CalibDrift}. As $\Delta a_{max}$ increases, the quantity of FN decreases, while FP increases. 

In ARDENT, we opted to prioritise minimising the FN over the FP. Indeed, FN are systems wrongly classified as unstable, leading to their rejection. The resulting risk is to over-estimate the dynamical constraints, excluding potential valid zones of the period-mass space. On the other side, a large amount of FP leads to poorer constraints on the dynamical detection limits, because many systems are wrongly kept after applying the stability constraint. While this needs to be avoided, it does not lead to wrong results as opposed to large FN numbers. 
Imposing a rate of FN smaller than 5$\%$, and minimising the rate of FP accordingly, we converge to a drift threshold $\Delta a_{max}$=0.025$\%$. It is represented by the diamond symbol in Fig. \ref{Fig:CalibDrift}. At this stability threshold, the rate of FP is $\sim 50\%$. This is a conservative approach that favours the reliability of the results over their ability to refine the updated mass limits. We recommend this strategy, and define the default drift threshold in ARDENT as $\Delta a_{max}$=0.025$\%$. However, we emphasise that the $\Delta a_{max}$ parameter in ARDENT can be easily modified by the user (via the argument \texttt{max\_drift\_a} in the function \texttt{ARDENT\_DetectionLimitStab}). Importantly, the calibration experiments presented here assumed a certain system architecture, restricting their validity. The optimal stability threshold may differ for significantly different system architectures. The best approach consists in testing various drift thresholds or calibrating this drift again, when ARDENT is to be employed on systems with different architectures than the CJ-SP case presented here.

\FloatBarrier

\section{ARDENT command lines for TOI-1736} \label{App:CodeTOI1736}
This section illustrates ARDENT practically, and details the command lines used to produce the results presented in Sect. \ref{Sect:TOI-1736} on TOI-1736. The residual SOPHIE RVs, obtained after removing the best fit model, were derived prior to run this code. These residuals, together with an exhaustive overview of ARDENT functions, are available at \url{https://github.com/manustalport/ardent}. 
Noteworthy, the functions computing the data-driven and dynamical detection limits are distinct. As a result, it is possible to skip the computation of data-driven detection limits if they are already available, and to derive the dynamical constraints straight away. We refer to the link above for further details. 

\begin{figure*}
    \centering
\includegraphics[width=\textwidth]{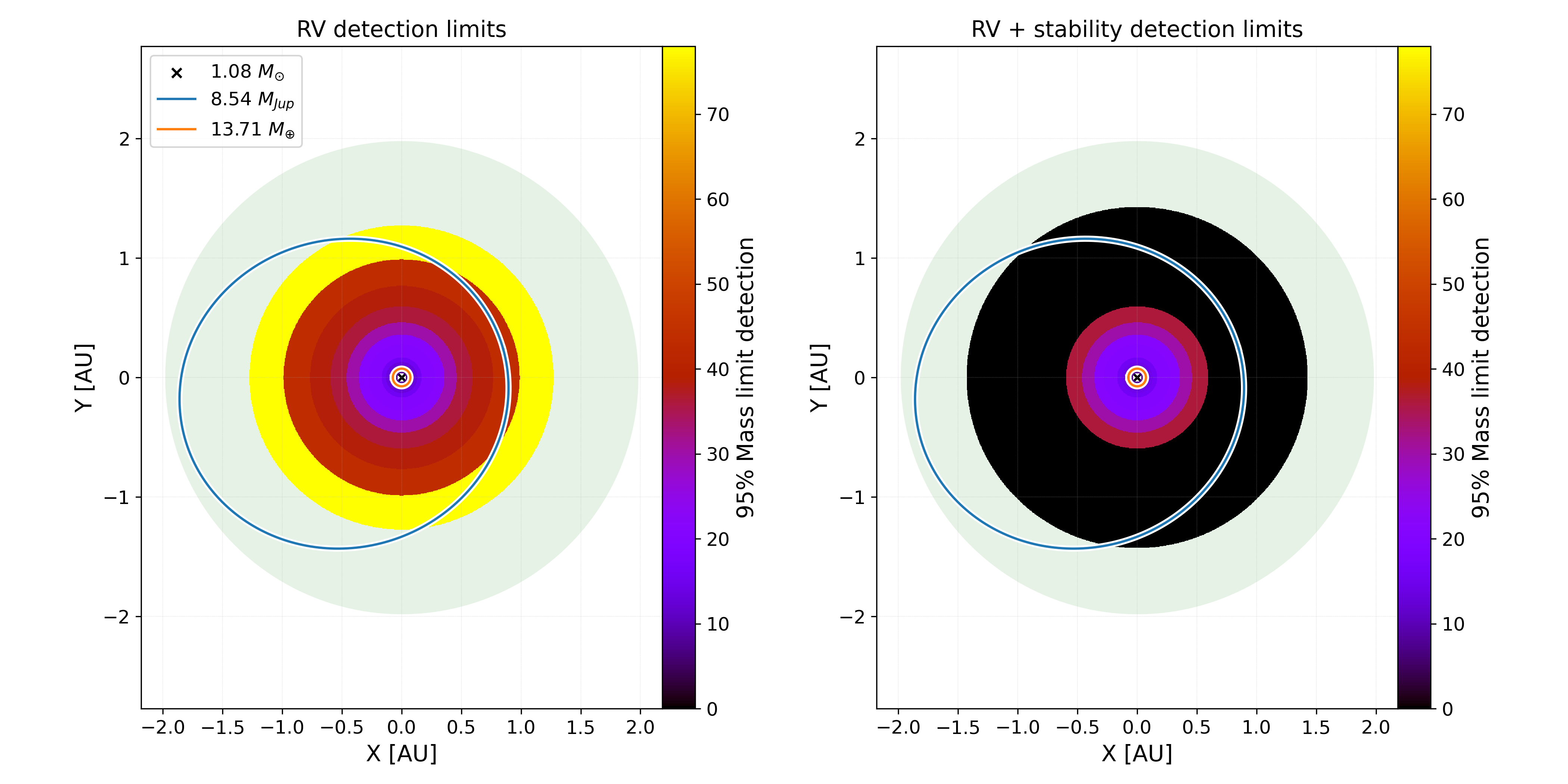}
\caption{TOI-1736 detection limits presented in the orbital plane. The star and the two known planetary orbits are represented. The green shaded region is the predictive habitable zone, whose extent is computed from \citet{Kopparapu2013}. The concentric discs illustrate the detection limits, with a colour code depicting the 95$\%$ limit mass. \textit{Left}: Data-driven detection limits, based on the SOPHIE RVs only. \textit{Right}: Dynamical detection limits, which add the constraint of orbital stability.} 
\label{Fig:DetectLim_OrbitalPlane} 
\end{figure*}


\begin{verbatim} 
import ardent 
import numpy as np 

data = np.genfromtxt( \
    'TOI1736_sophie_residualRV.dat', \
    skip_header=int(2))
jdb, rv, rv_err = data[:,0:3].T

# Initialise ardent
vec = ardent.ARDENT_tableXY(jdb, rv, rv_err) 

# Define an output folder that will contain the 
# results of this ardent run
vec.ARDENT_Set_output_dir('TOI-1736')

# Initialise the star and planets
vec.ARDENT_AddStar(mass=1.08,starname='TOI-1736')
vec.ARDENT_AddPlanets(p=570.3, k=200.4, \
    mean_long=244.0, e=0.362, omega=164.3, \
    asc_node=0.0, inc=70.0)
vec.ARDENT_AddPlanets(p=7.065, k=4.34, \
    mean_long=247.2, e=0.21, elower=0.0, \
    omega=0.0, asc_node=0.0, inc=90.0)

# Data-driven detection limits
vec.ARDENT_DetectionLimitRV(rangeP=[2., 600.]) 

# Dynamical detection limits
vec.ARDENT_DetectionLimitStab(NlocalCPU=4)

# Plot the detection limits
vec.ARDENT_Plot_StabDL(inset_plot=True, \
    zoom_xlim=[5.6,9.0], zoom_ylim=[0,16])

# Plot these limits in the orbital plane
vec.ARDENT_FinalPlot()
\end{verbatim}

The last line above produces an alternative representation of the detection limits in the orbital plane. It is presented in Fig. \ref{Fig:DetectLim_OrbitalPlane}, where the planetary orbits are drawn on top of the detection limits. This is visually informative to measure the impact of the stability constraints on the detection limits, and sense the extent of the zone where potential unseen planets could exist \citep[e.g.][]{Brewer2020}. The left plot shows the data-driven detection limits, to be compared with the plot on the right presenting the dynamical detection limits. These limits were computed in a period range between 2 and 600 days; hence,  the coloured areas do not extend until the borders of the plots. The impact of the eccentric giant planet is very effective in pushing down the detection limits. Any follow-up RV survey must focus their search for additional planets in the inner part of the system only, within the red and blue zones of the plot on the right.

\end{appendix}

\end{document}